\documentclass[12pt]{article}
\usepackage[dvips]{graphicx}
\title{Charmless two-body B decays: \\
A global analysis with QCD factorization}
\author{Dongsheng Du$^{a}$,
        Junfeng Sun$^{a}$,    Deshan Yang$^{a}$
        and Guohuai Zhu$^{b}$ \\
{\small \em a. Institute of High Energy Physics,
               Chinese Academy of Sciences,} \\
{\small \em  P.O.Box 918(4),
             Beijing 100039, China}
 \thanks{E-mail: duds@mail.ihep.ac.cn, sunjf@mail.ihep.ac.cn,
  yangds@mail.ihep.ac.cn.} \\
{\small \em b. Theory Group, KEK, Tsukuba, Ibaraki 305-0801, Japan}
 \thanks{E-mail: zhugh@post.kek.jp.}}
\begin{document}
\maketitle
\begin{abstract}
In this paper, we perform a global analysis of $B \to PP$ and $PV$ decays 
with the QCD factorization approach. It is encouraging to observe that the
predictions of QCD factorization are in good agreement with experiment. 
The best fit $\gamma$ is around $79^\circ$. 
The penguin-to-tree ratio $|P_{\pi \pi}/T_{\pi \pi}|$ of $\pi^+ \pi^-$ 
decays is preferred to be larger than $0.3$. We also show the confidence
levels for some interesting channels: $B^0 \to \pi^0 \pi^0$, $K^+ K^-$ 
and $B^+ \to \omega \pi^+$, $\omega K^+$. For $B \to \pi K^\ast$ decays,
they are expected to have smaller branching ratios with more precise 
measurements.
\end{abstract}
\newpage
\section{INTRODUCTION}

The charmless two-body B decays play a crucial role in determining the flavor 
parameters, especially the Cabibbo-Kobayashi-Maskawa (CKM) angles $\gamma$ 
and $\alpha$. With the 
successful running of B factories, many charmless decay channels have been 
measured with great precision. However, since hadronic B decays involve 
three separate scales, $m_W$, $m_b$, and $\Lambda_{QCD}$, where perturbative 
and nonperturbative effects are entangled, it is highly nontrivial to 
relate flavor parameters to experimental observables.

Recently, theorists have made much progress in nonleptonic B decays: 
three novel methods, QCD factorization (QCDF)\cite{QCDF}, the perturbative QCD 
approach (pQCD)\cite{pQCD} and the charming penguin method\cite{charming}, 
have been proposed. These methods have very different understandings of B 
decays: For both the QCDF and pQCD approaches, the factorization theorem is proved 
for nonleptonic B decays in the leading power expansion, i.e., short-distance 
physics related to the scales $M_W$ and $m_b$ can be separated from 
long-distance physics related to the hadronization scale $\Lambda_{QCD}$, 
and the long distance part can be parameterized into some universal 
nonperturbative parameters. In this sense, they are similar. But the pQCD approach 
implements the Sudakov form factor to suppress the end-point contributions 
and proves the factorization theorem in which the form factors are 
perturbatively calculable. Notice that Sudakov form factor itself is a 
perturbative quantity; it is rather radical and controversial to prove the 
factorization using the Sudakov form factor, while in QCDF the form factors 
are believed to be nonperturbative parameters. Therefore these two 
methods have completely different power behaviors for B decays. Their
predictions of B decays are also quite different. For instance, pQCD
generally predicts large strong phases and direct CP violations, while
QCDF favors small direct CP violations in general because of the
$\alpha_s$-suppressed strong phases. The charming penguin process, i.e., 
$(b \bar{q}) \to (c \bar{q})(\bar{c} s) \to (q^\prime \bar{q})
(\bar{q}^\prime s)$, might be potentially important for penguin-dominant
decays because it is doubly enhanced by CKM factors and Wilson
coefficients. The characteristic of the charming penguin method is that the
soft-dominance charming penguin plays an indispensable role for 
penguin-dominant decays. While in QCDF charm penguin contributions are hard
dominance and therefore perturbatively calculable 
according to naive power counting rules. 

Now BaBar and Belle have accumulated copious data, and will record much
more data, on nonleptonic B decays. Thus it should be highly interesting
to compare the predictions of these methods with precise experimental
measurements. We gave the QCDF predictions on B ${\to}$ PP and PV decays
in recent works \cite{ourPP,ourPV}. With the experimental data at that
time, our results prefer a somewhat larger angle $\gamma$. For PV decays,
the QCDF predictions are only marginally consistent with the experimental
observation for some decay channels. Notice that the QCDF predictions
contain large numerical uncertainties due to the CKM matrix elements, form
factors, and annihilation parameters, and furthermore, the uncertainties of
various decay channels are strongly correlated to each other; we are
stimulated to do a global analysis in this paper to check the consistency
between the predictions of QCDF and the updated experimental results.
Beneke {\it et al.} \cite{BBNS1} have done a global analysis including
$\pi \pi$, $\pi K$ modes with the QCDF approach and have shown a satisfactory
agreement between the QCDF predictions and experiments, while in this
work we shall consider not only $B \to PP$ decays, but also $B \to PV$
channels. Thereby, as we will see later, it leads to some new interesting
results.

One of the most impressive predictions of the QCDF approach is that 
direct CP violation of charmless B decays should be small because the
strong interaction phase arises solely from radiative corrections.
Up to now it has been very consistent with the measurements of BaBar and Belle.
However, power corrections which may also contribute to strong phases
are numerically comparable with radiative corrections. Notice that the
power corrections are difficult to estimate because they generally break
factorization. It means that the predictions of QCDF on direct CP violations
are probably qualitative. Therefore in this paper we will not consider 
experimental results on direct CP violations.

Our global fit shows that QCDF has an excellent performance on $B \to
PP$ (two light pseudoscalars) decays except for the channel $B^+ \to \eta
K^+$. But we do not worry about it because of the hard-to-estimate
contributions from the digluon mechanism and the potential large power
corrections in this channel. The CKM angle $\gamma$ is preferred to be 
around $79^\circ$ which is slightly larger but still consistent with the
standard CKM global analysis\cite{global}. We also discuss the preferred
range of the penguin-to-tree ratio $|P_{\pi \pi}/T_{\pi \pi}|$ 
\footnote{The defination of the penguin-to-tree ratio 
$|P_{{\pi}{\pi}}/T_{{\pi}{\pi}}|$ for $B_{d}^{0}{\to}{\pi}^{+}{\pi}^{-}$
decay is \cite{BBNS1}
\[ {\cal A}(B_{d}{\to}{\pi}^{+}{\pi}^{-})\ {\propto}\ \ e^{-i{\gamma}}
  + \frac{P_{{\pi}{\pi}}}{T_{{\pi}{\pi}}}, \]
\[ \frac{P_{{\pi}{\pi}}}{T_{{\pi}{\pi}}} = \frac{-1}{R_{b}}
\frac{(a_{4}^{c}+r^{\pi}_{\chi}a_{6}^{c}) 
    +(a_{10}^{c}+r^{\pi}_{\chi}a_{8}^{c})    
    +r_{A}[b_{3}+2b_{4}-\frac{1}{2}(b_{3}^{EW}-b_{4}^{EW})]}
{(a_{1}+a_{4}^{u}+r^{\pi}_{\chi}a_{6}^{u})
          +(a_{10}^{u}+r^{\pi}_{\chi}a_{8}^{u})
    +r_{A}[b_{1}+b_{3}+2b_{4}-\frac{1}{2}(b_{3}^{EW}-b_{4}^{EW})]}, \]
where $R_{b}=\frac{|V_{ud}V_{ub}^{\ast}|}{|V_{cd}V_{cb}^{\ast}|}
=\sqrt{{\bar{\rho}}^{2}+{\bar{\eta}}^{2}}$, and
$r_{A}{\simeq}\frac{f_{B}f_{\pi}}{m_{B}^{2}F_{0}^{B{\to}{\pi}}}$. } which is 
crucial for the extraction of angle $\alpha$. 
For
$B \to PV$ (one pseudoscalar, one vector) decays, QCDF has also a good
performance where the annihilation topology plays an important role
especially for penguin-dominated decays. But for $B \to \pi K^\ast$
channels, the QCDF results seem smaller compared with the experimental
measurements. However, presently there are large experimental errors on these
channels, so it would be very interesting for BaBar and Belle to update
their measurements with higher precision on these decay modes. Based on the
global fit, we also give the confidence levels for some interesting
decay channels: $B^0 \to \pi^0 \pi^0$, $K^+ K^-$, and $B^+ \to \omega
\pi^+$, $\omega K^+$.

This paper is organized as follows: in Sec. II, we will first
recapitulate the mainpoint of QCD factorization for charmless two-body B
decays. In Sec. III, the relevant input parameters are discussed. Then
the numerical results of the global fit and brief remarks are presented in
Sec. IV. Section V is devoted to the conclusions.

\section{QCD FACTORIZATION FOR CHARMLESS B DECAYS}

As we know, charmless B decays contain three distinct scales: 
$M_W \gg m_b \gg \Lambda_{QCD}$. To go beyond the naive model estimation,
it is important to show that the physics of different scales can be
separated from each other. This process is generally called 
``factorization'' .

It is well known that, with the help of the operator product expansion and
renormalization group equation, the effective Lagrangian can be obtained, in
which short-distance effects involving large virtual momenta of the
loop corrections from the scale $M_W$ down to $\mu={\cal O}(m_b)$ are
cleanly integrated into the Wilson coefficients. Then the amplitude for
the decay $B \rightarrow M_1 M_2$ can be expressed as \cite{Buras}
\begin{equation}
{\cal A}(B \rightarrow M_1 M_2) = \frac{G_F}{\sqrt{2}}
\sum_i \sum_{q=u,c} \lambda_q C_i(\mu)
\langle M_1 M_2 \vert Q_i(\mu) \vert B \rangle,
\end{equation}
where $\lambda_q$ is a CKM factor, $C_i(\mu)$ is the Wilson coefficient
which is perturbatively calculable from first principles, and 
$\langle M_1 M_2 \vert Q_i(\mu) \vert B \rangle $ 
is a hadronic matrix element which contains physics from the scale
$\mu={\cal O}(m_b)$ down to $\Lambda_{QCD}$. In a sense, this process 
may be called ``first step factorization''. But it is still highly
nontrivial to estimate the hadronic matrix elements reliably because 
the perturbative and nonperturbative effects related to $m_b$ and
$\Lambda_{QCD}$ are strongly entangled.

Three years ago, Beneke {\it et al.} put forward the 
QCDF approach in the heavy quark limit for $B \to \pi \pi$ \cite{QCDF}.
They show that, neglecting power corrections in $1/m_b$, the hadronic 
matrix elements can be factorized into hard radiative corrections and a 
nonperturbative part parameterized by the form factors and meson light
cone distribution amplitudes. In the following we will outline their
reasoning.

\subsection{QCDF in the heavy quark limit}
First, we need to have some knowledge about the end-point behavior of the
light cone distribution amplitudes (LCDAs) of the mesons. At the scale of
$m_b$, the LCDAs of the final light mesons---for example ${\phi}(x)$ of
$\pi$ or $K$ mesons---should be similar to the asymptotic form. Therefore
it is reasonable to assume that the end-point of the LCDAs of the light
mesons is suppressed by $\Lambda/m_b$. For B mesons, the spectator quark is
assumed to be soft and have no hard tail, i.e., $\phi(\xi) \sim m_b/\Lambda$,
for $\xi < \Lambda/m_b$ and $\phi(\xi)=0$, for $\xi > \Lambda/m_b$.
With the above assumptions, the form factor is argued to be nonperturbative
dominant; thereafter, naive power counting rules are constructed and the
leading power radiative contributions in $1/m_b$ can be identified (see
Fig. 1).
\begin{figure}[htb]
\vspace*{-0.5cm}
\centerline{\includegraphics[height=8cm,width=18cm]{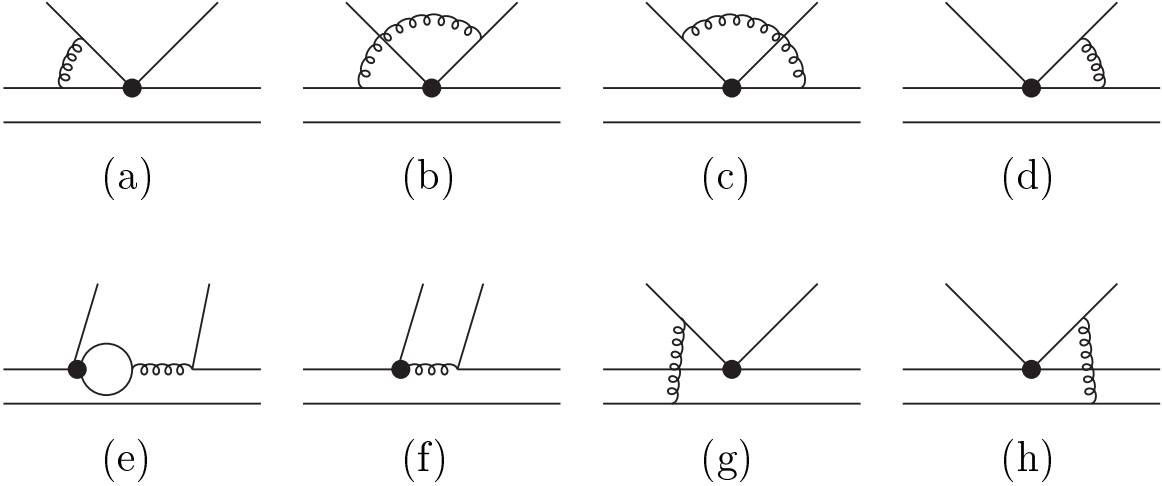} }
\vspace*{-0.5truecm}
\caption{Order of $\alpha_s$ corrections to the hard scattering kernels.
     The upward quark lines represent the emission meson from the b quark 
     decay vertex. These diagrams are commonly called vertex corrections,
     penguin corrections and hard spectator scattering diagrams for
     (a)-(d), (e) and (f), and (g) and (h), respectively.}
\end{figure}

Notice that, in Fig. 1, the emission meson from the decay vertex carries large
energy and momentum (about $m_B/2$) and therefore can be described by
leading twist-2 LCDA in the leading power approximation. For factorization
to be held, these radiative contributions should be hard dominant. For
vertex corrections (Figs. 1(a)-1(d)), every individual diagram contains
infrared divergence, but these infrared divergences are canceled after
summation. This cancellation is not accidental. Intuitively, the $q \bar{q}$
pair of the energenic emission meson can be viewed as a small color
dipole. Since soft gluons can not taste the difference between a small color 
dipole and a color singlet, the 
emission meson decouples with the soft gluon interaction. This argument is
well known as ``color transparency'' \cite{Bjorken}. Technically, not only
soft divergence but also collinear divergence is canceled. For penguin
corrections (Figs. 1(e) and 1(f)) and hard spectator scattering 
(Figs. 1(g) and 1(h)), 
since the end point of the twist-2 LCDA of the light meson is $\Lambda/m_b$
suppressed, it is not difficult to show hard dominance. So factorization
does hold in the heavy quark limit, and the corresponding formula can be
explicitly expressed as
\begin{eqnarray}
\langle M_1 M_2 \vert Q_i \vert B \rangle &=&
F^{B \rightarrow M_2}(0) \int \limits_0^1 dx T^I_{i1}(x) \Phi_{M_1}(x)
+F^{B \rightarrow M_1}(0) \int \limits_0^1 dx T^I_{i2}(y) \Phi_{M_2}(y)
\nonumber \\
& &+\int \limits_0^1 d\xi dx dy T_i^{II}(\xi,x,y) \Phi_B(\xi) \Phi_{M_1}(x)
\Phi_{M_2}(y) \nonumber \\
&=&\langle M_1 M_2 \vert J_1 \otimes J_2 \vert B \rangle
\cdot [1+ \sum r_n \alpha_s^n +{\cal O}(\Lambda_{QCD}/m_b)].
\end{eqnarray}
In the above formula, $\Phi_B(\xi)$ and $\Phi_{M_i}(x)(i=1,2)$ are the
leading twist wave functions of B and the light mesons, respectively, and
$T^{I,II}_i$ denote hard scattering kernels which are perturbatively
calculable. The readers may refer to Ref.\cite{BBNS2} for more details.

One of the most interesting results of the QCDF approach is that, in the
heavy quark limit, strong phases are short dominant and arise solely from
vertex and penguin corrections which are at the order of $\alpha_s$. It
means that, for charmless hadronic decays, direct CP violations are
generally small because strong phases are $\alpha_s$ suppressed compared
to the leading ``naive factorization'' contributions. But in principle
power corrections may also contribute to strong phases, and numerically
$\Lambda_{QCD}/m_b$ is comparable to $\alpha_s$. Furthermore, there is
no known systematic way to estimate power suppressed contributions (note
that soft collinear effective theory \cite{Bauer} may be a potential
tool), so QCDF could only predict strong phases qualitatively.

\subsection{Chirally enhanced power corrections}
The above discussions are based on the heavy quark limit; i.e., power 
corrections in $1/m_b$ are assumed to be negligible. Then the question 
is, for phenomenological applications, whether it is a good approximation.
There are various sources which may contribute to power corrections in
$1/m_b$; examples are higher twist distribution amplitudes, transverse
momenta of quarks in the light meson, annihilation diagrams, etc. 
At first sight, power corrections seem really small because they 
are suppressed by $\Lambda_{QCD}/m_b \simeq 1/15$. However, this is not
true. For instance, the contributions of operator $Q_6$ to decay 
amplitudes would formally vanish in the strict heavy quark limit. But 
it is numerically very important in penguin-dominated B rare decays, such
as the interesting $B \rightarrow \pi K$ decays. This is because
$Q_6$ is always multiplied by a formally power-suppressed but
chirally enhanced factor 
$r_{\chi}=\frac{2 m_{P}^2}{m_b(m_1+m_2)} \sim {\cal O}(1)$, where $m_1$
and $m_2$ are current quark masses. 
Another example is annihilation topology (Fig. 2), the importance of which was 
noticed first in the pQCD method \cite{pQCD}.
Therefore phenomenological applicability of QCD factorization in B rare
decays requires at least a consistent inclusion of chirally enhanced
corrections and annihilation contributions.
  \begin{figure}[htb]
  \vspace*{-0.cm}
  \centerline{\includegraphics[height=4cm,width=16cm]{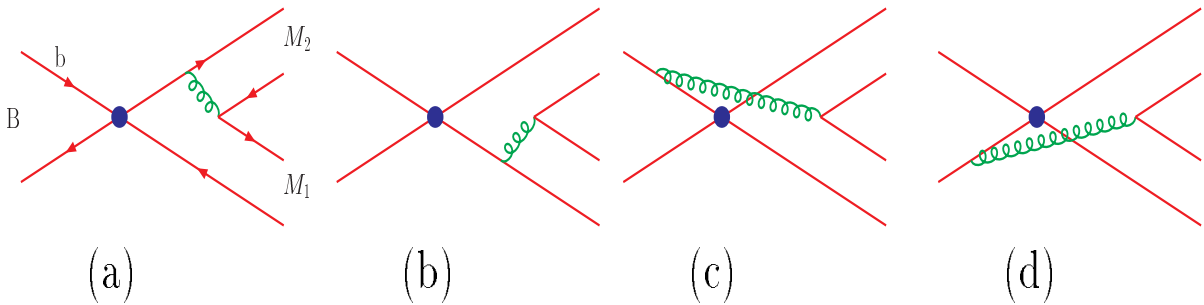} }
  \vspace*{-0.truecm}
  \caption{Order of $\alpha_s$ corrections to the weak annihilations.}
  \end{figure}

Chirally enhanced corrections arise from twist-3 light cone distribution
amplitudes; thus, the final light mesons should be described by leading
twist and twist-3 distribution amplitudes. Then we need to redemonstrate
that the leading power radiative corrections (Fig. 1) are still dominated
by hard gluon exchange. Unfortunately it is not true for hard spectator
scattering which contains logarithmic divergence in the end-point region.
A similar divergence also appears in the annihilation contributions. 
It means that, strictly speaking, factorization does not hold for chirally
enhanced corrections and annihilation topology. The readers may refer to
Refs. \cite{BBNS1,DYZ} for more technical details. Phenomenologically,
Beneke {\it et al.} \cite{BBNS1} introduced a model parametrization for 
the end-point divergence:
\begin{equation}
X_{A,H}=\int \limits_0^1 \frac{dx}{x} = \ln \frac{m_B}{\Lambda_{h}}
(1+\rho_{A,H}\,e^{i\phi_{A,H}}),
\end{equation}
where $X_A$ denotes the annihilation contribution and $X_H$ denotes hard
spectator scattering. We will follow their approach in this paper.

For the rest of the power corrections, they are argued to be generally small
\cite{Neubert} based on a model estimation with renormalon calculus.

With the above discussions, the decay amplitudes can be written as
\begin{equation}
{\cal A}(B \to M_1 M_2)=\frac{G_F}{\sqrt{2}}\sum_{p=u,c} \sum_{i=1}^{10}
v_p ( a_i^p \langle M_1 M_2 \vert O_i \vert B \rangle_f +
f_B f_{M_1} f_{M_2} b_i ),
\end{equation}
where $\langle M_1 M_2 \vert O_i \vert B \rangle_f$ is the factorized
hadronic matrix element which has the same definition as that in the 
naive factorization approach. For the explicit expressions of QCD
coefficients $a_i$ and annihilation parameters $b_i$, the readers may
refer to Refs. \cite{ourPP,ourPV,BBNS1}
\footnote{In fact, there is minor difference for the hard spectator 
scattering term $f^{II}$ between ref.\cite{ourPP} and ref.\cite{BBNS1}. This is 
due to a subtle point relating to some divergent boundary terms in a process of 
integration by part \cite{Benekekek}. In this paper, we adopt the expression of 
$f^{II}$ from ref.\cite{BBNS1}. }.

\section{INPUT PARAMETERS}

The decay amplitude for $B \to M_1 M_2$ depends  on various parameters,
such as the CKM matrix elements, decay constants, form factors,
renormalization scale $\mu$, LCDAs, and so on. Notice that although the predictions
of QCDF are formally scale independent at one-loop order, numerically 
there is still a small residual dependence. In the global fit, the scale
$\mu$ is varied from $m_b/2$ to $2 m_b$. For the rest of the parameters,
we will specify them in the following.

\subsection{CKM matrix elements}
The CKM matrix in the Wolfenstein parametrization is read as
 \begin{equation}
 V_{CKM}=\left(\begin{array}{ccc}
     1-\lambda^2/2  &    \lambda      &   A\lambda^3(\rho-i\eta)  \\
      -\lambda      &  1-\lambda^2/2  &   A\lambda^2              \\
      A\lambda^3(1-\rho-i\eta)  &  -A\lambda^2  &  1
         \end{array}\right) +{\cal O}(\lambda^4).
 \end{equation}
It contains four parameters $A$, $\lambda$, $\rho$ and $\eta$, in which
the first two are well determined \cite{hocker}:
\[ \lambda=|V_{us}|=0.2200\pm0.0025, \hspace{1cm}
  A\lambda^2=|V_{cb}|=(40.4\pm1.3_{stat}\pm0.9_{theo})\times 10^{-3}.
\]
As to $\rho$ and $\eta$, they are kept free except for the constraint
$|V_{ub}|=(3.49 \pm 0.24_{stat} \pm 0.55_{theo}) \times 10^{-3}$
\cite{hocker} and $\sin 2\beta= 0.731 \pm 0.055$ (world averaged)
\cite{2beta}.

\subsection{Form factors and decay constants}

The form factors and decay constants are nonperturbative parameters. 
The form factors can be extracted from the semileptonic decays and/or 
estimated with some well-defined theories, such as lattice
calculations, QCD sum rules, etc. But the related errors are still 
sizable. The decay constants can be extracted from the leptonic or
electromagnetic decay width with high precision. In the fit, we choose 
the corresponding numerical values as follows
\cite{Ali,Neubert1,Feldmann,Ball}:
 \begin{eqnarray*}
 & &f_{\pi}=131\mbox{~MeV},    \ \ \ \ \ \
    f_{K}=160\mbox{~MeV},      \ \ \ \ \ \
    f_{K^{\ast}}=214\mbox{~MeV}, \\
 & &f_{\rho}=210\mbox{~MeV},   \ \ \ \ \ \
    f_{\omega}=195\mbox{~MeV}, \ \ \ \ \ \ \
    f_{\phi}=233\mbox{~MeV},     \\
 & &f_{q}=140\mbox{~MeV}, \ \ \ \ \ \
    f_{s}=176 \mbox{~MeV}, \ \ \ \ \ \
    \phi=39.3^\circ,   \\
 & &f_B=(180\pm 40)\mbox{~MeV}, \ \ \ \ \ \ \ \ \ \ \ \ 
    R_{\pi K}=0.9 \pm 0.1, \\
 & &F^{B{\pi}}_{0,1}(0)=0.28{\pm}0.05, \ \ \ \ \ \ \ \ \ \ \ \
    A^{BK^{\ast}}_{0}(0)=0.47{\pm}0.07,\\
 & &A^{B{\rho}}_{0}(0)=0.37{\pm}0.06,  \ \ \ \ \ \ \ \ \ \ \ \
    A^{B{\omega}}_{0}(0)=A^{B\rho}_0 (0),
 \end{eqnarray*}
where $R_{\pi K} \equiv f_\pi F^{BK}/f_K F^{B\pi}$. In the above, we assume 
ideal mixing between $\omega$ and $\phi$, i.e., 
$\omega=(u\bar u +d\bar d)/\sqrt{2}$ and $\phi=s\bar s$. 
As for $\eta-\eta^{\prime}$ mixing, we follow the convention in the
quark-flavor basis \cite{Leutwyler,Kroll} and assume that the charm quark
content in $\eta^{(\prime)}$ is negligible,
 \begin{eqnarray}
 & &{\langle}0{\vert}\bar{q}{\gamma}_{\mu}{\gamma}_{5}q
           {\vert}{\eta}^{(\prime)}(p){\rangle}
          =if^{q}_{{\eta}^{(\prime)}}p_{\mu}
    \ \ \ \ \ \ \ (q=u,d,s), \nonumber \\
 & &\frac{{\langle}0{\vert}\bar{u}{\gamma}_{5}u
                    {\vert}{\eta}^{(\prime)}{\rangle}}
         {{\langle}0{\vert}\bar{s}{\gamma}_{5}s
                    {\vert}{\eta}^{(\prime)}{\rangle}}
  = \frac{f^{u}_{{\eta}^{(\prime)}}}{f^{s}_{{\eta}^{(\prime)}}},
    \ \ \ \ \ \
          {\langle}0{\vert}\bar{s}{\gamma}_{5}s
                    {\vert}{\eta}^{(\prime)}{\rangle}
  = -i\frac{m_{{\eta}^{(\prime)}}^{2}}{2m_{s}}
   (f^{s}_{{\eta}^{(\prime)}}-f^{u}_{{\eta}^{(\prime)}}), \nonumber \\
 & &f^{u}_{\eta}=  \frac{f_{8}}{\sqrt{6}}{\cos}{\theta}_{8}
                -  \frac{f_{0}}{\sqrt{3}}{\sin}{\theta}_{0},
    \ \ \ \ \ \ \ \
    f^{s}_{\eta}=-2\frac{f_{8}}{\sqrt{6}}{\cos}{\theta}_{8}
                -  \frac{f_{0}}{\sqrt{3}}{\sin}{\theta}_{0},\\
 & &f^{u}_{{\eta}^{\prime}}=  \frac{f_{8}}{\sqrt{6}}{\sin}{\theta}_{8}
                           +  \frac{f_{0}}{\sqrt{3}}{\cos}{\theta}_{0},
    \ \ \ \ \ \ \ \
    f^{s}_{{\eta}^{\prime}}=-2\frac{f_{8}}{\sqrt{6}}{\sin}{\theta}_{8}
                           +  \frac{f_{0}}{\sqrt{3}}{\cos}{\theta}_{0},
    \nonumber \\
 & &F_{0,1}^{B{\eta}}=F_{0,1}^{B{\pi}}\Big(
         \frac{{\cos}{\theta}_{8}}{\sqrt{6}}
        -\frac{{\sin}{\theta}_{0}}{\sqrt{3}}\Big), \ \ \ \ \
    F_{0,1}^{B{\eta}^{\prime}}=F_{0,1}^{B{\pi}}\Big(
         \frac{{\sin}{\theta}_{8}}{\sqrt{6}}
        +\frac{{\cos}{\theta}_{0}}{\sqrt{3}}\Big), \nonumber
 \end{eqnarray}
where the four octet-singlet parameters can be related to three
quark-flavor parameters:
\begin{eqnarray}
& &f_8=\sqrt{1/3 f_q^2 + 2/3 f_s^2}, \ \ \ \ \ \ \ \ \ \ \ \
f_0=\sqrt{2/3 f_q^2 + 1/3 f_s^2}, \nonumber \\
& &\theta_8=\phi-\arctan{(\sqrt{2}f_s/f_q)}, \ \ \ \ \ \
\theta_0=\phi-\arctan{(\sqrt{2}f_q/f_s)}.
\end{eqnarray}

\subsection{LCDAs of the mesons}

The LCDAs of the mesons are basic input parameters in the QCDF approach.
The LCDAs of a light pseudoscalar meson are defined as \cite{LCDA,Beneke}
\begin{eqnarray}
\langle P(p') \vert {\bar q_{\alpha}}(y) q_{\delta}(x) \vert 0 \rangle
&=&\frac{if_P}{4} \int_0^1{\it du~e}^{i(up' \cdot y + {\bar u}p' \cdot
x)} \nonumber \\
&\times& \left \{ \slash{\hskip -2.5mm}p^{\prime} \gamma_5 \phi(u)
-\mu_P \gamma_5 \left ( \phi_p(u)-\sigma_{\mu \nu}p^{\prime \mu}
z^{\nu} \frac{\phi_{\sigma}(u)}{6} \right ) \right \}_{\delta \alpha},
\end{eqnarray}
where $z=y-x$, $\phi(u)$ ($\phi_{p,\sigma}(u)$) is leading twist (twist-3)
LCDA, and $\mu_P=m_P^2/(m_1(\mu)+m_2(\mu))$ (here $m_1$ and $m_2$ are
current masses of the valence quarks of the pseudoscalar meson). Because
the current masses of light quarks are difficult to fix, we would like to
take
\[ 
r_{\eta}\left ( 1-\frac{f_{\eta}^u}{f_{\eta}^s} \right )= r_{\pi} = r_{K}=r_\chi ,
\]
which is numerically a good approximation. For the related quark masses,
we shall follow Ref. \cite{BBNS1}:
\[ 
m_s(2\mbox{~GeV})=(110\pm 25)\mbox{ MeV}, \hspace{0.8cm}
m_c(m_b)=1.3\mbox{ GeV}, \hspace{0.8cm} m_b(m_b)=4.2\mbox{ GeV}.
\]

For vector mesons, only longitudinal polarization is involved in 
$B \to PV$ decays. Furthermore, the contributions of twist-3 LCDAs of
vector mesons are doubly suppressed by $\alpha_s$ and $\Lambda/m_b$; 
therefore, they can be safely disregarded. Then the leading twist LCDA 
of a longitudinal vector meson is defined as \cite{LCDA,Beneke}
\begin{equation}
\langle V_\Vert(p') \vert {\bar q_{\alpha}}(y) q_{\delta}(x) \vert 0
\rangle = \frac{f_V m_V}{4} \int_0^1 {\it du} \, 
 e^{i(up' \cdot y + {\bar u}p' \cdot x)} \phi_\Vert(u) 
\slash{\hskip -2.5mm} p^{\prime}_{\delta \alpha}/E \, .
\end{equation}

We shall use the asymptotic forms of the LCDAs for the following
discussions:
\begin{equation}
\phi(u)=\phi_\Vert(u)=6u \bar{u}, \hspace{1cm}
\phi_p(u)=1, \hspace{1cm} \phi_\sigma(u)=6u \bar{u} \, .
\end{equation}
Strictly speaking, the asymptotic forms are only valid for 
$\mu \to \infty$. We notice that in Ref. \cite{BBNS1}, Beneke {\it et al.}
employ an expansion in Gegenbauer polynomials for leading twist $\pi, K$
LCDAs. However, since there are many light mesons involved in our global
fit, if we consider a similar expansion for the leading twist LCDAs, many
free parameters would be introduced. Fortunately, the corrections to the
asymptotic form are numerically not so important because they only affect 
part of the vertex and penguin corrections (numerically, the readers may 
refer to Table 3 of Ref. \cite{BBNS1} to see the effects of the Gegenbauer 
expansion). So for simplification,
only the asymptotic forms are used in our discussions.

For the wave function of the B meson, only the moment
$\int_0^1 d\xi \Phi_B(\xi)/\xi \equiv m_B/\lambda_B$ appears in the
factorization formulas. We do not know much about the parameter
$\lambda_B$ and the estimation of Ref. \cite{BBNS1} is quoted :
$\lambda_B = (350 \pm 150)\mbox{ MeV}$.

\section{GLOBAL ANALYSIS OF CHARMLESS B DECAYS}

In this work, the global analysis is based on the CKMFitter package
\footnote{http://ckmfitter.in2p3.fr}
developed by H\"ocker {\it et al.} \cite{global}. The original package includes
$B \to h h$ ($h=\pi$ or $K$) decay channels, and we enlarge it to include
$B \to PV$ and $B \to \eta \pi (K)$ decay modes. The Rfit scheme is 
implemented for statistical treatment. Simply speaking, the Rfit scheme
assumes the experimental errors to be pure Gaussians (if the systematic
errors are not so large) while the theoretical parameters vary freely in 
a given range. In this spirit, it is similar to the $95\%$ scan method. One
of the main differences is that the overall $\chi^2_{min}$ is assumed to
be Gaussian distributed in the $95\%$ scan method, while for the Rfit scheme, the
confidence level of the overal $\chi^2_{min}$ is computed by means of a 
Monte Carlo simulation. The readers may refer to Ref. \cite{global} for
details about the Rfit scheme. As to the QCDF expressions for the related
decay amplitudes, the readers may refer to Refs. \cite{BBNS1,ourPP}
for B $\to$ PP decays and \cite{ourPV} for B $\to$ PV decays.

Compared with pQCD, QCDF requires more input parameters, such as form
factors, annihilation parameter $X_A$,  hard spectator parameter $X_H$,
and so on. To make the global analysis appear more persuasive and at the
same time save computing time, we minimize the number of variables by
fixing those insensitive parameters. One example is that we shall use the
asymptotic forms of the LCDAs for all final light mesons, and not employ
an expansion in Gegenbauer polynomials.

Since power corrections violate factorization, the parameters $X_A$ and
$X_H$ are introduced as a model parameterization. But we should be care
that, in principle, these parameters are channel dependent. Fortunately,
assuming ``factorized'' SU(3) breaking, we can see that $X_A$ and $X_H$
are universal separately for $B \to PP$ and $B \to PV$ decay modes.
However, there is no way to relate the chiral parameters of the $PV$
channels to those of the $PP$ channels. So we have to introduce, besides 
$X_A^{PP}$ and $X_H^{PP}$, the additional
parameters $X_A^{PV}$ and $X_H^{PV}$ for $B \to PV$ decays.

In Refs. \cite{BBNS1,Beneke1}, Beneke {\it et al.} present a detailed
analysis of $B \to \pi\pi$, $\pi K$ with the QCDF approach. They show an
impressive agreement between experiments and the QCDF predictions: 
$\chi^2 \approx 0.5$ for six decay channels. Their best fit results favor
$\gamma$ around $90^\circ$ which seems not so consistent with the standard
global fit of the CKM matrix elements using information from semileptonic 
B decays, $K$-$\bar K$ mixing and $B$-$\bar B$ mixing. Even for $\gamma$
around $60^\circ$, $\chi^2 \approx 1$ is still good enough to be
acceptable. In QCDF, these six decay channels are sensitive to several
input parameters: the CKM parameters $|V_{ub}|$ and angle $\gamma$ (or
equivalently $\rho$ and $\eta$), form factors $F^{B\pi}$ and $F^{BK}$,
annihilation-related parameters $X_A$ and $f_B/\lambda_B$, and current
quark mass $m_s$. These parameters vary freely only in a given range
which is either determined by experimental measurements ($|V_{ub}|$),
estimations with QCD sum rules and/or lattice calculations (form factors
and decay constants), or well-educated guesswork ($X_A$). So it is really
nontrivial for the achieved agreement between the QCDF predictions and
the experimental measurements.

In this work, we will extend the global analysis to include 14 
$B \to PP$ and $PV$ decay modes (see Table 1). Notice that the hard spectator
parameter $X_H$ is numerically unimportant for the branching ratios
except for $a_2$-related tree-dominated decays \cite{ourPV}, and even for
$a_2$-related decays, it brings at most $20\%$ uncertainties. So compared
with the global analysis of $B \to \pi \pi$, $\pi K$ \cite{BBNS1}, our
extension would include seven $B \to PV$ channels and newly observed
$B \to \eta \pi$ decay, while only three new sensitive parameters --- the form
factor $A_0^{B\rho}$ and complex variable $X_A^{PV}$ --- would be 
involved. Therefore we can have a more stringent test of the QCDF 
predictions which should give some interesting information.

Recently BaBar and Belle also gave a strong constraint on direct CP
violations for many charmless hadronic B decay channels. Their search 
show that direct CP-violating asymmetries are generally small. Within 
the QCDF framework, strong phases are either $\alpha_s$ or $\Lambda/m_b$
suppressed, which also lead to small direct CP violations in general.
However, only radiative corrections are perturbatively computable in QCDF,
while power corrections break the factorization in general. Considering
that $\Lambda/m_b$ is numerically comparable with $\alpha_s$, the QCDF
calculations on direct CP violations are probably qualitative. So the
experimental constraints on direct CP violations are not implemented in
this global analysis.

Before doing the global fit, the readers may notice that some decay modes
are not included in the global analysis although they have been observed.
These decay channels are listed in Table 2, and we will discuss these
channels later.

\begin{table}[htb]
\begin{center}
\caption{Experimental data of CP-averaged branching ratios for
some charmless B decay modes in units of $10^{-6}$. The following
decay channels are the experimental input of the global fit.}
\begin{tabular}{lcccc}\hline \hline
BF($\times 10^6$)& CLEO \cite{CLEO}& BaBar \cite{BaBar} &
Belle \cite{Belle} & Average  \\ \hline
$B^0 \to \pi^+ \pi^-$ & $4.3^{+1.6}_{-1.4}\pm 0.5 $ &
$4.7 \pm 0.6 \pm 0.2$ & $5.4 \pm 1.2 \pm 0.5      $ &
$4.77 \pm 0.54     $
\\
$B^+ \to \pi^+ \pi^0$ & $5.4^{+2.1}_{-2.0}\pm 1.5 $ &
$5.5^{+1.0}_{-0.9}\pm 0.6$ & $7.4^{+2.3}_{-2.2} \pm 0.9      $ &
$5.78 \pm 0.95        $
\\
$B^0 \to K^+ \pi^-  $ & $17.2^{+2.5}_{-2.4}\pm 1.2$ &
$17.9 \pm 0.9\pm 0.7$ & $22.5 \pm 1.9 \pm 1.8     $ &
$18.5 \pm 1.0       $
\\
$B^+ \to K^+ \pi^0  $ & $11.6^{+3.0}_{-2.7}~^{+1.4}_{-1.3}$ &
$12.8^{+1.2}_{-1.1}\pm 1.0$ & $13.0^{+2.5}_{-2.4} \pm 1.3     $ &
$12.7 \pm 1.2       $
\\
$B^+ \to K^0 \pi^+  $ & $18.2^{+4.6}_{-4.0}\pm 1.6$ &
$17.5^{+1.8}_{-1.7}\pm 1.3$ & $19.4^{+3.1}_{-3.0} \pm 1.6     $ &
$18.1 \pm 1.7       $
\\
$B^0 \to K^0 \pi^0  $ & $14.6^{+5.9}_{-5.1}~^{+2.4}_{-3.3}$ &
$10.4 \pm 1.5 \pm 0.8$ & $8.0^{+3.3}_{-3.1} \pm 1.6      $ &
$10.2 \pm 1.5        $
\\
$B^+ \to \eta \pi^+ $ & $<5.7                    $ &
$<5.2               $ & $5.3^{+2.0}_{-1.7}(<8.2) $ &
$<5.2               $
\\
$B^0 \to \pi^\pm \rho^\mp$ & $27.6^{+8.4}_{-7.4}\pm 4.2 $ &
$28.9 \pm 5.4 \pm 4.3$     & $20.8^{+6.0}_{-6.3}~^{+2.8}_{-3.1} $ &
$25.4 \pm 4.3      $
\\
$B^+ \to \pi^+ \rho^0$ & $10.4^{+3.3}_{-3.4}\pm 2.1 $ &
$<39                 $ & $8.0^{+2.3}_{-2.0}\pm 0.7 $ &
$8.6 \pm 2.0        $
\\
$B^0 \to K^+ \rho^-  $ & $16.0^{+7.6}_{-6.4}\pm 2.8        $ &
$                    $ & $11.2^{+5.9}_{-5.6}~^{+1.9}_{-1.8}$ &
$13.1 \pm 4.7        $
\\
$B^+ \to \phi K^+        $ & $ 5.5^{+2.1}_{-1.8}\pm 0.6$ &
$9.2 \pm 1.0 \pm 0.8$ & $10.7 \pm 1.0^{+0.9}_{-1.6}$ &
$8.9 \pm 1.0             $
\\
$B^0 \to \phi K^0        $ & $5.4^{+3.7}_{-2.7}\pm 0.7$ &
$8.7^{+1.7}_{-1.5}\pm 0.9$ & $10.0^{+1.9}_{-1.7}~^{+0.9}_{-1.3}$ &
$8.6 \pm 1.3             $
\\ 
$B^+ \to \eta \rho^+  $ & $<10 $ &
$                     $ & $<6.2$ &
$<6.2                 $
\\
$B^0 \to \omega K^0 $ & $ <21  $ &
$5.9^{+1.7}_{-1.5}\pm 0.9$ & $                            $ &
$5.9 \pm 1.9            $
\\ \hline \hline
\end{tabular}
\end{center}
\end{table}
\begin{table}[htb]
\begin{center}
\caption{Measurements which are not included in the global analysis.}
\begin{tabular}{lcccc}\hline \hline
BF($\times 10^6$)& CLEO \cite{CLEO} & BaBar \cite{BaBar} &
 Belle \cite{Belle} & Average  \\ \hline
$B^+ \to \eta K^+   $ & $<6.9                     $ &
$<6.4               $ & $5.2^{+1.7}_{-1.5}(<7.7)  $ &
$<6.4               $
\\ 
$B^+ \to \pi^+ K^{\ast 0}$ & $ <16                      $ &
$15.5\pm 3.4 \pm 1.8     $ & $16.2^{+4.1}_{-3.8}\pm 2.4 $ &
$15.8 \pm 3.0        $
\\ 
$B^0 \to \pi^- K^{\ast +}$ & $  16^{+6}_{-5} \pm 2   $ &
$                        $ & $26.0 \pm 8.3 \pm 3.5        $ &
$19.0 \pm 4.9            $
\\ 
$B^+ \to \eta K^{\ast +}   $ & $26.4^{+9.6}_{-8.2}\pm 3.3   $ &
$22.1^{+11.1}_{-9.2}\pm 3.3$ & $26.5^{+7.8}_{-7.0}\pm 3.0   $ &
$25.4 \pm 5.3              $
\\ 
$B^0 \to \eta K^{\ast 0}  $ & $13.8^{+5.5}_{-4.6}\pm 1.6   $ &
$19.8^{+6.5}_{-5.6}\pm 1.7$ & $16.5^{+4.6}_{-4.2}\pm 1.2   $ &
$16.4 \pm 3.0             $
\\ 
$B^+ \to \omega K^+  $ & $<8   $ &
$<4                  $ & $9.2^{+2.6}_{-2.3}\pm 1.0   $ &
$                    $
\\ 
$B^+ \to \omega \pi^+    $ & $11.3^{+3.3}_{-2.9}\pm 1.5   $ &
$6.6^{+2.1}_{-1.8}\pm 0.7$ & $<8.2                        $ &
$                        $
\\ \hline \hline
\end{tabular}
\end{center}
\end{table}

\subsection{Main results of the global fit}
When the decay channels in Table 1 are concerned, the global fit shows
that the QCDF predictions are well consistent with the experimental
measurements: The results in the $(\overline{\rho}, \overline{\eta})$ 
plane are shown in Fig. 3 where 
$\chi^2_{min} = 4.2$ for $14$ decay channels. As an
illustration, in Table 3, we list the best fit values of the global
analysis for the related $B \to PP$,$PV$ decay modes with and without
chiral-related contributions. Notice that two sets of best fit values
(with or without chirally enhanced contributions) are obtained with 
different input parameters. It indicates that the newly observed 
$B^0 \to \omega K^0$ decay can be included in the global fit without any
difficulty, and that it is hopeful that the decay $B^+ \to \eta \rho^+$ will be
observed soon. The corresponding theoretical inputs for the best fit values 
including chirally enhanced corrections are also reasonable: 
$\vert V_{ub} \vert = 3.57 \times 10^{-3}$, $\gamma=79^\circ$, 
$F^{B\pi}=0.24$, $A_0^{B\rho}=0.31$, $m_s=85~\mbox{MeV}$, 
$\mu=2.5~\mbox{GeV}$, $f_B=220~\mbox{MeV}$, 
$\rho_A^{PP}=0.5$, $\phi_A^{PP}=10^\circ$, $\rho_A^{PV}=1$, 
$\phi_A^{PV}=-30^\circ$. As to $F^{BK}$, there is no strong constraint 
and the range $[0.24,0.30]$ is acceptable from the 
current global analysis. Since the hard spectator contributions are nearly 
negligible except for $a_2$-related tree-dominant decays
($B \to \pi^+ \pi^0, \pi^+ \rho^0$), and even for $a_2$-related tree decays, 
it brings at most $20\%$ uncertainties to the branching ratios \cite{ourPV}, 
the global 
analysis can not give a strong constraint on the hard spectator parameter $X_H$.  
In principle, the chirally enhanced corrections could 
lead to large strong phases from the imaginary part of the annihilation 
topologies. But since the best fit parameters show a small imaginary part: 
$\phi_A^{PP}=10^\circ$,  $\phi_A^{PV}=-30^\circ$; the global analysis still 
prefers small direct CP violations, which is consistent with the current 
experimental observations. 
In Ref.
\cite{Beneke1}, it is argued that chirally enhanced
corrections are not indispensable for $B \to \pi \pi$ and $\pi K$ decays.
However, we can see from Table 3 that, especially for penguin-dominated
$B \to PV$ decays, chirally enhanced contributions play an important role.
Note that this point is not firmly established: There are significant
experimental errors in $B^0 \to K^+ \rho^-$, $\omega K^0$ decays. If
these two channels were excluded, we could see from Table 3 that it is still
acceptable without chiral-related contributions. However, 
$\pi^+ K^{\ast 0}$ decay also implies large chiral contributions:
Without chirally enhanced corrections, it is clear that
\begin{equation}
\frac{{\cal A}(B^+ \to \pi^+ K^{\ast 0})}
{{\cal A}(B^+ \to \pi^+ K^0)} \simeq \frac{f_{K^\ast}}{f_K}
\frac{a_4}{a_4+a_6} \simeq 1/2 \, ,
\end{equation}
which means
\begin{equation}
 {\cal B}(B^+ \to \pi^+ K^{\ast 0})
\simeq \frac{1}{4}{\cal B}(B^+ \to \pi^+ K^0) \simeq 5 \times
10^{-6} \, .
\end{equation}
It is 3 times smaller than the experimentally central value (see 
Table 2). Thereby further measurements with higher precision on the
penguin-dominated $B \to PV$ decays will clarify the role of the chirally
enhanced contributions. We will return back to $\pi K^\ast$ channel later
and explain why we do not include this mode in the global fit.

\begin{table}[htb]
\begin{center}
\caption{The best fit values using the global analysis with and without 
chiral-related contributions for $B \to PP$ and $PV$ decays. ``No chiral''
means the best fit value neglecting the chirally enhanced hard spectator
contributions and the annihilation topology. The branching ratios are in
units of $10^{-6}$. The experimental data are the uncorrelated average of
measurements of BaBar, Belle, and CLEO (see the data in the last column of
Table 1).}
\begin{tabular}{lccccc} \hline \hline
Mode & $B^0 \to \pi^+ \pi^-$ & $B^+ \to \pi^+ \pi^0$ &
$B^0 \to K^+ \pi^-$ & $B^+ \to K^+ \pi^0$ & $B^+ \to K^0 \pi^+$ \\ \hline
Expt. & $4.77 \pm 0.54$ & $5.78 \pm 0.95$ &
$18.5 \pm 1.0$ & $12.7 \pm 1.2$ & $18.1 \pm 1.7$ \\ 
Best fit  & $4.82$ & $5.35$ &
$19.0$ & $11.4$ & $20.1$ \\ 
No chiral & $5.68$ & $3.25$ & $18.8$ & $12.6$ & $20.2$ \\ \hline
Mode & $B^0 \to \pi^0 K^0$ & $B^+ \to \eta \pi^+$ &
$B^0 \to \rho^\pm \pi^\mp$ & $B^+ \to \rho^0 \pi^+$ &
$B^+ \to \eta \rho^+$ \\ 
Expt. & $10.2 \pm 1.5$ & $<5.2$ &
$25.4 \pm 4.3$ & $8.6 \pm 2.0$ & $<6.2$ \\ 
Best fit  & $8.2$ & $2.8$ &
$26.7$ & $8.9$ & $4.6$ \\ 
No chiral & $7.3$ & $1.8$ & $29.5$ & $8.5$ & $3.8$ \\ \hline
Mode & $B^+ \to \phi K^+$ & $B^0 \to \phi K^0$ &
$B^0 \to K^+ \rho^-$ & $B^0 \to \omega K^0$ \\ 
Expt. & $8.9 \pm 1.0$ & $8.6 \pm 1.3$ &
$13.1 \pm 4.7$ & $5.9 \pm 1.9$ \\ 
Best fit  & $8.9$ & $8.4$ &
$12.1$ & $6.3$ \\ 
No chiral & $7.1$ & $6.7$ &
$ 5.1$ & $1.2$ \\ \hline \hline
\end{tabular}
\end{center}
\end{table}

It is known that the penguin-to-tree ratio $|P_{\pi \pi}/T_{\pi \pi}|$
is very useful for extraction of the CKM angle $\alpha$ \cite{Gronau}. 
In Ref. \cite{BBNS1}, the authors show that 
$|P_{\pi \pi}/T_{\pi \pi}|=(28.5 \pm 5.1 \mp 5.7)\% $ 
with the QCDF approach using the default values for the chirally enhanced
corrections, i.e., $X_{H}=X_{A}=\ln (m_B/\Lambda_h)$. When considering the
uncertainties of the $X_{A,H}$ parameters, the theorectical errors would
be even larger. So it should be very interesting to obtain the preferred
$|P/T|$ ratio \footnote{The decay amplitudes fot $B_d^0 \to \pi^+ \pi^-$ are 
\cite{Gronau}
 \[ {\cal A}(B_d \to \pi^+ \pi^-) = -( |T| e^{i \delta_T} e^{i \gamma}
   + |P| e^{\delta_P} ), \]
 where $\delta_T$ and $\delta_P$ are strong phases.}
from the global analysis. Until now the asymptotic LCDAs were used 
for the global fit because the branching ratios are numerically not so
sensitive to the corrections to the asymptotic form. But for the 
penguin-to-tree ratio, the case is different and we should consider the
Gegenbauer polynomial expansion  for the leading twist LCDAs of the
$\pi$, $K$ mesons. We find that, compared with the estimation 
$|P/T| = 0.276 \pm 0.064$ \cite{Gronau} (including $SU(3)$ breaking
effects), the global fit prefers a suprisingly large value: 
$|P_{\pi \pi}/T_{\pi \pi}|\simeq 0.41$. The reason may be that penguin
annihilation effects increase the penguin amplitudes, as discussed in
\cite{BBNS1}. Considering the relatively large model dependence of this 
ratio within the QCDF framework, the best fit value may be not so
meaningful to reduce the ambiguity in the determination of $\sin 2\alpha$.
However, even assuming that it is acceptable for the global fit with
$\chi^2/N_{dof} \leq 1$ ($N_{dof}$ denotes the number of degrees), the
ratio $|P_{\pi \pi}/T_{\pi \pi}|$ is still larger than $0.3$. This result
is quite interesting, although undoubtedly it needs further tests with
larger data samples.

The confidence levels of the angle $\gamma$ and some interesting decay channels 
are given in Fig. 3 and 4, respectively. It is encouraging that the
favored angle $\gamma$ is around $79^\circ$ which is somewhat larger but
still consistent with the standard CKM global fit. 
${\cal B}(B^0 \to \pi^0 \pi^0)$, which is crucial for a clean extraction
of the angle $\alpha$, is predicted to be around $1 \times 10^{-6}$. So it is
hopeful to be observed in the near future. For the pure annihilation decay
$B^0 \to K^+ K^-$, although the updated upper limit has been very
stringent, the branching ratio is predicted to be still several times
smaller than that: roughly $10^{-7}$. 

It is interesting to have a brief look at the relevant PQCD results. Based on 
$\pi \pi$, $\pi K$ decays, PQCD could extract the central values of the CKM 
angles \cite{Keum}: $\alpha=78^\circ$, $\beta=26^\circ$, $\gamma=76^\circ$. These 
results are (probably coincident) consistent with our best fit results. But 
${\cal B}(B^0 \to \pi^0 \pi^0)$ is about $0.3 \times 10^{-6}$ in the PQCD method, 
which is smaller than the QCDF prediction and could be tested soon in the near 
future. PQCD also predicts a large direct CP violation for $\pi^+ \pi^-$ decay. 
\begin{figure}[htb]
\centerline{\includegraphics[height=7.3cm,width=16cm]{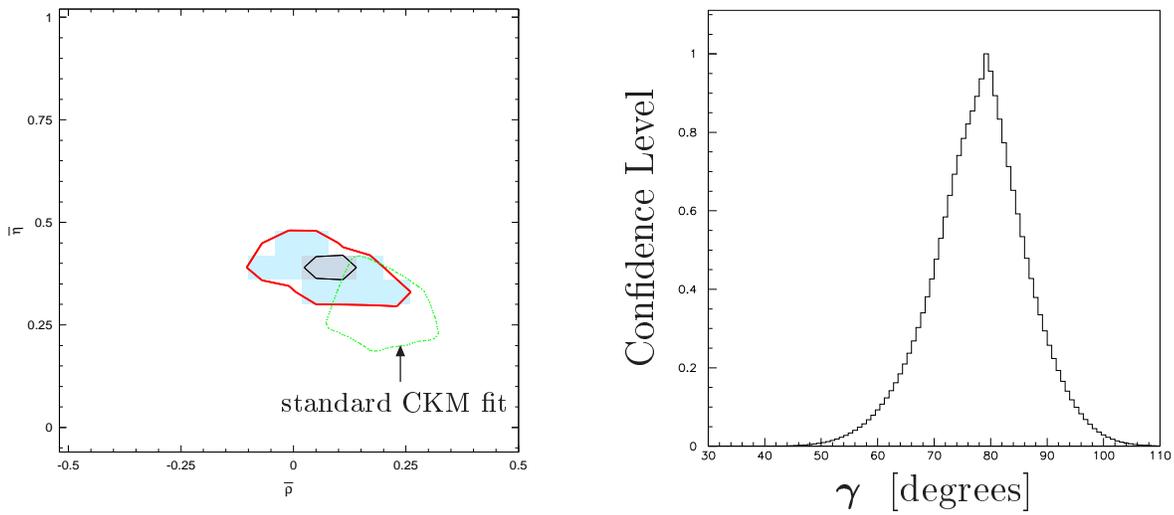} }
\caption{Left plot:  
confidence level in the $(\overline \rho, \overline \eta)$ 
plane for the global fit. The contours with shaded area inside indicate the 
regions of $\ge 90 \% $ and $\ge 5 \% $ Confidence levels. 
Right plot:  
the confidence level for angle $\gamma$ in units of degree.}
\end{figure}

\begin{figure}[htb]
\centerline{\includegraphics[height=20cm,width=14cm]{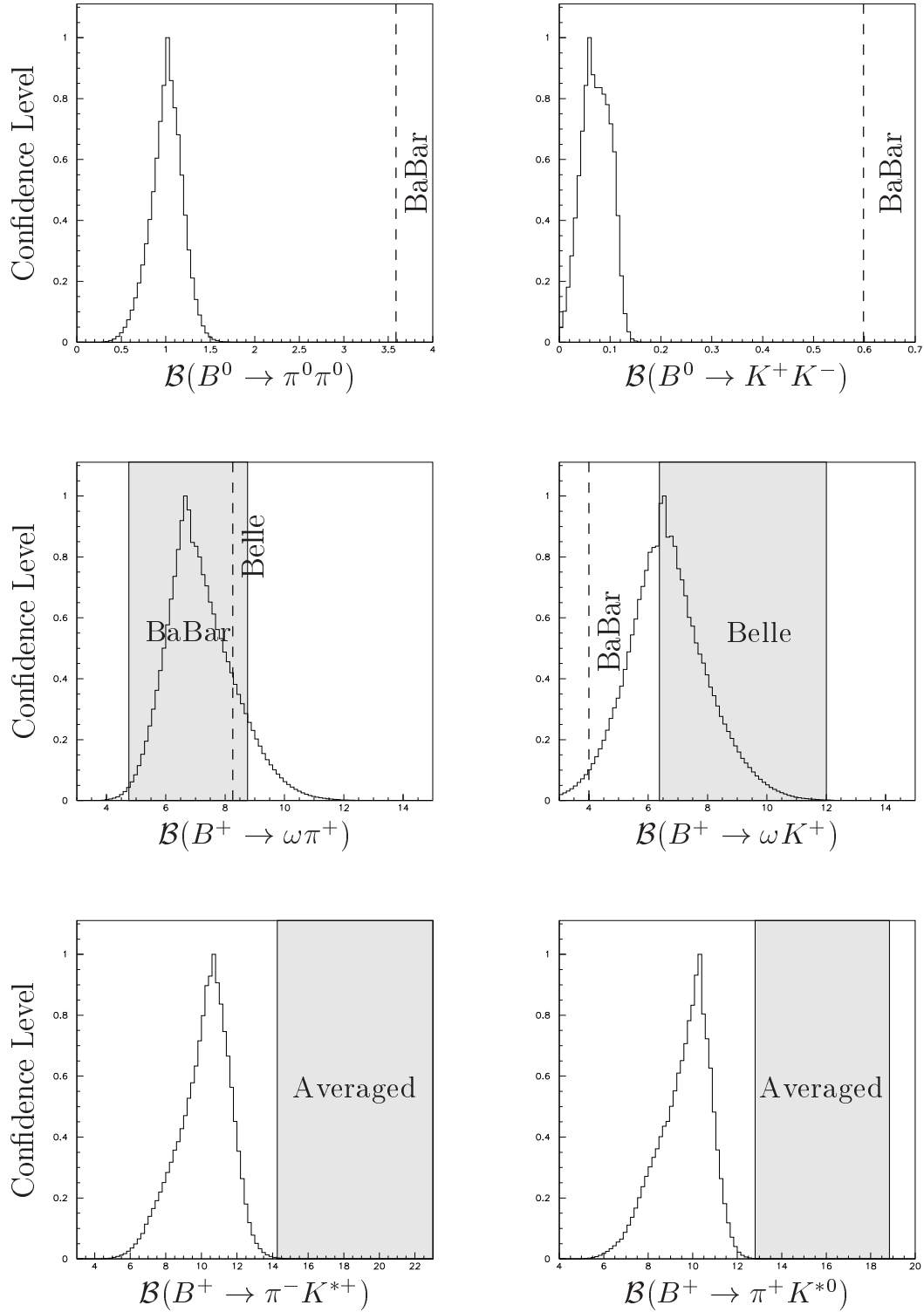} }
\caption{The confidence levels for some selected decay channels. The branching 
ratios are in units of $10^{-6}$. The dashed lines denote the experimental 
upper limits; the gray bands denote the experimental measurements with 
$1\sigma$ error.} 
\end{figure}
\clearpage

\subsection{Decay modes listed in Table 2}

Now let us discuss in detail why we do not include the decay modes listed 
in Table 2 for the global analysis. 

For $B^+ \to \omega \pi^+ (K^+)$, the measurements of Belle are not so
consistent with those of BaBar and CLEO. Assuming 
$A_0^{B\omega} = A_0^{B\rho}$ which should be a good approximation, the 
confidence levels for these two decay channels are shown in Fig. 4. For
$\omega \pi^+$ decay, the best fit value is $6.66 \times 10^{-6}$ which is
consistent with the measurements of BaBar and Belle. The best fit value
for $\omega K^+$ decay is $6.25 \times 10^{-6}$ which is consistent with
the Belle observation but larger than the upper limit given by BaBar. 

As to $B \to \eta K^\ast$ decays, the branching ratios depend on the form
factor $A_0^{BK^\ast}$. Note that
there are no other observed $PV$ decay
channels relying on this form factor; it is more or less trivial to 
include $B \to \eta K^\ast$ decays, since in some sense $A_0^{BK^\ast}$ 
acts essentially as a free parameter in the global fit especially 
considering the large experimental errors in these decay channels. One
might argue that the observed $B \to \phi K^\ast$ decay also depends on
the form factor $A_0^{BK^\ast}$. But this decay channel requires a new
annihilation parameter $X_A^{VV}$ for $B \to VV$ decays. Hence it may be
better to first have a restricted constraint on $X_A^{VV}$ with more 
$B \to VV$ decay modes to be observed. Then more precise data on 
$B \to \phi K^\ast$, $\eta K^\ast$ could overconstrain the form factor
$A_0^{BK^\ast}$ and give a more stringent test of the QCDF approach.

The decay mode $B^+ \to \eta K^+$ was recently observed by Belle. Within
the QCDF framework, the corresponding amplitude is proportional to
$f_K F^{B\eta}(a_4+r_\chi a_6)+f_\eta^s F^{BK}(a_4+ r_\chi a_6)$. Notice
that $f_\eta^s <0$, and large cancellation occurs which leads to a much smaller
branching ratio compared with the experimental data. But we do not worry
about it due to several reasons.
\begin{itemize}
\item First, it relates to the special property of $\eta^{(\prime)}$
which has anomaly coupling to two gluons. Specifically, the digluon fusion
mechanism \cite{digluon} where one gluon comes from the $b \to s$ decay 
vertex and the other from the spectator quark, is presumed to account for the
large branching ratios of $B \to \eta^\prime K$ decays. Although it is
arguable whether the digluon mechanism is perturbatively calculable
or not, the contribution should be proportional to the coupling
$\langle 0 \vert G \tilde{G} \vert \eta^{(\prime)} \rangle$.
As we know that
\[
\frac{\langle 0 \vert \frac{\alpha_s}{4\pi}G \tilde{G} \vert \eta
\rangle}{\langle 0 \vert \frac{\alpha_s}{4\pi}G \tilde{G} \vert
\eta^\prime \rangle} =\left ( \frac{M_\eta}{M_{\eta^\prime}}
\right )^2 \cot \phi \simeq \frac{1}{5}\, \mbox{~~~and~~~}
\left \vert \frac{ {\cal A}^{\mbox{expt}}(B\to \eta K) }
{ {\cal A}^{\mbox{expt}}(B\to \eta^\prime K) } \right \vert \simeq
\frac{1}{4} \, ,
\]
this means that 
\[
\left \vert \frac{ {\cal A}^{g}(B\to \eta K) }
{ {\cal A}^{\mbox{expt}}(B\to \eta K) } \right \vert
\sim \left \vert \frac{ {\cal A}^{g}(B\to \eta^\prime K) }
{ {\cal A}^{\mbox{expt}}(B\to \eta^\prime K) } \right \vert ,
\]
where $A^{g}$ denotes the amplitude of the digluon fusion mechanism. So if the
digluon mechanism were important for $B \to \eta^\prime K$, it should be
also important for $B \to \eta K$.
\item Second, we know that when the leading power terms are abnormally
small, the next-to-leading power contributions become potentially
important. Remembering that there is no known systematic way to estimate
power corrections, the QCDF estimation of this channel is probably correct
only at the order of magnitude.
\item Recently, Beneke \cite{Beneke1} propose a novel
possibility: for the annihilation contributions, two gluons may radiate
from the spectator quark and form a $\eta^{(\prime)}$ meson. In this case,
it is the leading power contribution. Furthermore, it breaks the factorization
and therefore a new nonperturbative parameter is needed to parameterize
its contribution. 
\end{itemize}
From the above discussions, it is clear that theoretically great efforts are
needed to quantitatively understand $B \to \eta^{(\prime)} K$ decays.

The real trouble is $B \to \pi K^\ast$ decays. Let us take 
$B^+ \to \pi^+ K^{\ast 0}$ as an example. In QCDF, approximately we have
\begin{eqnarray}
  &&{\cal A}(B^+ \to \pi^+ K^{\ast 0}) \propto a_4 f_{K^\ast} F^{B\pi}
 \times \mbox{const} + f_B f_{K^\ast} f_\pi b_3(V,P) \, ,\\
  &&{\cal A}(B^+ \to \phi K^+)  \propto a_4 f_\phi F^{BK} \times \mbox{const}
    + f_B f_\phi f_K b_3(V,P) \, .
\end{eqnarray}
Assuming $F^{BK}/F^{B\pi} \approx f_K/f_\pi$, then
\[
  \frac{ {\cal A}(B^+ \to \pi^+ K^{\ast 0}) }{{\cal A}(B^+ \to \phi K^+)}
  \approx \frac{f_{K^\ast} F^{B\pi}}{f_\phi F^{BK}} < 1 \, .
\]
So ${\cal B}(B^+ \to \pi^+ K^{\ast 0})$ should be smaller than or
at most comparable with ${\cal B}(B^+ \to \phi K^+)$. Unfortunately, 
the updated experimental measurements do not support it:
\[
 {\cal B}(B^+ \to \pi^+ K^{\ast 0}) = (15.8 \pm 3.0){\times} 10^{-6} \, ,
 \hspace{1.5cm}
 {\cal B}(B^+ \to \phi K^+) = (8.8 \pm 1.0){\times} 10^{-6} .
\]
General speaking, we do not anticipate any novel mechanism for this
channel because it would have a similar influence on $B \to \pi K$ decays.
So presently there is nothing we can do from the QCDF side. With the
global fit, the confidence level for $B \to \pi K^\ast$ is displayed in
Fig. 4, from which we can see that ${\cal B}(B^+ \to \pi^+ K^{\ast 0})$
is comparable with ${\cal B}(B^+ \to \phi K^+)$ as expected. The preferred
branching ratios are somewhat smaller than the experimental measurements
for both $\pi^+ K^{\ast 0}$ and $\pi^- K^{\ast +}$ modes. Fortunately the
current experimental errors are quite large. We anticipate 
that further precise measurements would prefer smaller branching ratios
for these two decay channels.

It is interesting to notice that, although 
there are essential difference between the PQCD and QCDF methods, 
PQCD also predicts 
that ${\cal B}(B^+ \to \pi^+ K^{\ast 0})$ is comparable with 
${\cal B}(B^+ \to \phi K^+)$ \cite{Keum1}, which is about $10 \times 10^{-6}$. 

\section{SUMMARY}
QCD factorization is a promising method for charmless two-body B 
decays, which are crucial for the determination of the unitarity triangle.
With the successful running of B factories, many $B \to PP$ and $PV$ decay
modes have been observed. We can do a global analysis to check whether
the predictions of the QCD factorization are consistent with
the measurements. Since there are many parameters involved in the global
analysis, we try to minimize the number of free QCD parameters to make
the global analysis appear more persuasive. Hence, the asymptotic forms
are used for the light cone distribution amplitudes of the light
pseudoscalar and vector mesons. For chirally enhanced parameters, it is a
good approximation to take $r_\eta=r_\pi=r_K=2m_K^2/m_b(m_s+m_u)$.
Assuming ``factorized'' SU(3) breaking, the chirally enhanced parameters
$X_{A,H}$ are separately universal for $B \to PP$ and $PV$ decays. 
However, $X_{A,H}^{PP}$ and $X_{A,H}^{PV}$ are independent parameters
because there is no approximate symmetry to relate $B \to PP$ and $B \to
PV$ decays.

With the above set of parameters, we enlarged the CKMFitter package to 
include more charmless decay channels and did a global analysis. It is
shown that the QCDF predictions are basically in good agreement with the
experiments. It is encouraging to see that the favored angle $\gamma$ is 
roughly consistent with the standard CKM global fit. It is quite
interesting  to see that the chirally enhanced corrections may play an
important role in penguin-dominated $B \to PV$ decays. The penguin-to-tree
ratio $\vert P_{\pi \pi}/T_{\pi \pi} \vert$ is important for the extraction 
of the CKM
angle $\alpha$. The global analysis favors this ratio to be larger than
$0.3$ with $\chi^2$ per number of degree smaller than 1. 

Notice that the observed decays $\eta K^+$, $\eta K^{\ast}$, $\pi K^\ast$,
$\omega \pi^+$, $\omega K^+$ are not included in the analysis. In this 
paper we discussed these decay channels in detail. Among these channels,
only the decays $B \to \pi K^\ast$ are somewhat troublesome. In QCDF,
${\cal B}(B^+ \to \pi^+ K^{\ast 0})$ should be smaller than or at most
comparable with ${\cal B}(B^+ \to \phi K^+$), which is not so consistent
with experimental observations. In fact, the global analysis prefers
somewhat smaller branching ratios for both $\pi^+ K^{\ast 0}$ and 
$\pi^- K^{\ast +}$ decay modes. Fortunately the related experimental
errors are quite large at present; it is anticipated that further
measurements with higher precision would observe smaller branching ratios. 
We also gave the confidence levels for some selected decay channels: 
$B^0 \to \pi^0 \pi^0$, $K^+ K^-$ and $B^+ \to \omega \pi^+(K^+)$.

\section*{ACKNOWLEDGMENTS}
We are grateful to S. Laplace for the help on the CKMFitter package.
G. Zhu also thanks A.I. Sanda, H.n. Li, and Y.Y. Keum for helpful
discussions. This work is supported in part by National Natural Science
Foundation of China. G. Z. thanks JSPS of Japan for financial support.

\end{document}